\documentclass[journal,twoside,web]{ieeecolor}
\usepackage{generic}

\usepackage{bm}
\usepackage{enumerate}
\usepackage{commath}
\usepackage{graphicx}
\graphicspath{{Pictures/}}
\usepackage{amsmath}
\usepackage{subcaption}
\usepackage{breqn}
\usepackage{cite}
\usepackage[table]{xcolor}
\usepackage{booktabs}
\usepackage{empheq}
\usepackage{amsfonts}
\usepackage{amssymb}

\usepackage{amsthm}
\usepackage{hyperref}
\usepackage{xcolor}
\usepackage{mathrsfs}
\usepackage{accents}
\usepackage{thmtools}
\usepackage{thm-restate}
\usepackage{float}
\usepackage{comment}
\usepackage{textcomp}

\theoremstyle{plain}
\newtheorem{lemma}{Lemma}
\newtheorem{definition}{Definition}
\newtheorem{assumption}{Assumption}

\newtheorem{proposition}{Proposition}
\newtheorem{remark}{Remark}

\newtheorem{problem}{Problem}
\newtheorem*{problem*}{Problem}
\newtheorem*{theorem*}{Theorem}
\newtheorem{assumption*}{Assumption}

\declaretheorem[name=Theorem]{thm}

\newcommand{\rev}[1]{{\color{black}#1}}

\newcommand{\myvar}[1]{\bm{#1}}

\newcommand{\barvar}[1]{\bar{\bm{#1}}}

\newcommand{\myvardot}[1]{\dot{\myvar{#1}}}

\newcommand{\myset}[1]{\mathcal{#1}}
\newcommand{\mysetbound}[1]{\partial \mathcal{#1}}
\newcommand{\mysetint}[1]{\text{Int}(\mathcal{#1})}

\newcommand{\robvar}[1]{\hat{\myvar{#1}}}

\newcommand{\mysetbar}[1]{\bar{\myset{#1}}}

\title{\LARGE \bf
On Compatibility and Region of Attraction for Safe, \\ Stabilizing Control Laws
}

\author{Wenceslao Shaw Cortez, \IEEEmembership{Member, IEEE} and Dimos V. Dimarogonas, \IEEEmembership{Senior Member, IEEE}
\thanks{This work was supported by the Swedish Research Council (VR),
the Swedish Foundation for Strategic Research (SSF), the Knut and Alice
Wallenberg Foundation (KAW) and the H2020-EU Research  and Innovation Programme under the GA No. 101016906 (CANOPIES).
The authors are with the School of EECS, Royal Institute of Technology (KTH), 100 44 Stockholm, Sweden (Email: 
        {\tt\small wencsc, dimos@kth.se}).}
}

\begin{document}

\maketitle
\thispagestyle{empty}
\pagestyle{empty}

\begin{abstract}
A novel method is proposed to ensure stability and constraint satisfaction, i.e. ``compatibility", for nonlinear affine systems. We require an asymptotically stabilizing control law and a zeroing control barrier function (ZCBF), and define a region of attraction for which the proposed control safely stabilizes the system. Our methodology requires checking conditions of the system dynamics over the state space, which may be computationally expensive. To facilitate the search for compatibility, we extend the results to a class of nonlinear systems including mechanical systems for which a novel controller is designed to guarantee passivity, safety, and stability. The proposed technique is demonstrated using numerical examples.
\end{abstract}

\section{Introduction}

Zeroing control barrier functions (ZCBFs) have gained attention for constraint satisfaction of nonlinear systems \cite{Ames2019}. ZCBFs are robust to perturbations \cite{Ames2019,Xu2015a}, can respect input constraints \cite{ShawCortez2020, Squires2018}, and are less restrictive than those of Lyapunov-based methods as the derivative of the ZCBFs need not be positive semi-definite \cite{Ames2019}. 

Other constraint satisfying methods include Nonlinear Model Predictive Control (NMPC) and reference governors (RG). ZCBFs are typically implemented in a one-step lookahead fashion that reduces the computational effort compared to NMPC where large prediction horizons may be required \cite{Grune2011}. RGs can be conservative as they require the Lyapunov function level sets to be completely contained in the constraint set \cite{Hosseinzadeh2020, Garone2017}. ZCBFs on the other hand allow the system to ``touch" the constraint boundary for less conservative behaviour \cite{Ames2019}. 

Despite the advantages of existing ZCBF methods, there is a lack of guarantees therein regarding simultaneous stability and constraint satisfaction (i.e safety), which we refer to as ``compatibility". In many cases it is desirable to know regions in which both the ZCBF and stabilizing conditions (i.e. decrescent Lyapunov function) hold simultaneously. For example, passivity is often coupled with a Lyapunov function (e.g., mechanical systems). Thus ensuring \rev{that} the Lyapunov function remains decrescent preserves passivity, which is favourable for interacting with the environment \cite{Schaft2017} and for interconnected systems. Earlier work considered ``safety-critical" ZCBF controllers \cite{Ames2017}, where constraint satisfaction was prioritized over stability. There have been extensions addressing event-triggering \cite{Yang2019}, sampled-data systems \cite{ShawCortez2019, Ghaffari2018}, high-relative degree \cite{Tan2021,Nguyen2016}, \rev{ZCBFs with NMPC \cite{Schilliger2021}}, and input constraints \cite{Zeng2021a, ShawCortez2020, Squires2018}, and \rev{learning \cite{Robey2020}}. Although safety is a priority, it is desirable to simultaneously ensure both constraint satisfaction and stability to safely accomplish a desired task. 

ZCBFs have been combined with a control Lyapunov function (CLF) in a CLF-ZCBF formulation to help stabilize the system \cite{Reis2021, Jankovic2018}. Local stability guarantees have been provided in \cite{Jankovic2018}, while \cite{Reis2021} eliminates undesirable equilibrium points, however there is no explicit region of attraction for which asymptotic stability holds. Furthermore, the CLF-ZCBF method uses a slack variable with respect to the stabilizing condition. Thus the given CLF may \emph{increase} in the safe set. Finally, those methods depend on a previously designed CLF, which for nonlinear systems is non-trivial to construct. The problem addressed here is to design a control law with an associated region of attraction that ensures \textit{compatibility} between a ZCBF and a nominal stabilizing control law that is not dependent on a CLF.

An important application for deriving such a region of attraction and an associated safe, stabilizing controller is planning and control of mechanical systems. In this application, the concept of safety is closely related to passivity to prevent an unknown environment (e.g. a human) from de-stabilizing the system \cite{Schaft2017}. Recent work relies on linearizing the system dynamics \cite{He2020} to conservatively address this problem. Another method employs a CLF-ZCBF controller, while being restricted to ellipsoidal constraints and assuming Lipschitz continuity of the control law \cite{Barbosa2020}. Furthermore, neither of those existing methods address passivity to properly address safety of the overall system.

In this paper, we address compatibility. The contribution is two-fold. First, we construct a region of attraction wherein safety and stability hold simultaneously. We design a novel control law that ensures asymptotic stability and safety over the region of attraction, which is not dependent on a CLF and allows Lyapunov functions that rely on LaSalle's invariance principle. Second, we extend our results to design a novel control law for mechanical systems that guarantees safety, stability, and passivity of the closed-loop system. Numerical simulations are used to validate the proposed method. 
 
\textit{Notation}: The Lie derivatives of a function $h(\myvar{x})$ for the system $\myvardot{x} = \myvar{f}(\myvar{x}) + \myvar{g}(\myvar{x}) \myvar{u}$ are denoted by $L_f h$ and $L_g h$, respectively. The term $\nabla h(\myvar{x})$ for a function $h: \mathbb{R}^n \to \mathbb{R}$ is the gradient of $h$ and $\nabla^2 h(\myvar{x})$ denotes its Hessian. The interior, boundary, and closure of a set $\myset{A}$ are denoted $\text{Int}(\myset{A})$, $\partial \myset{A}$, and $\text{clos}(\myset{A})$ respectively. The ball of radius $\delta$ about a point $\myvar{y}$ is defined by $\myset{B}_\delta(\myvar{y}) := \{ \myvar{x} \in \mathbb{R}^n: ||\myvar{x} - \myvar{y}|| \leq \delta \}$, and denoted by $\myset{B}_\delta$ if $\myvar{y} = 0$. The norm $||\myvar{y} ||^2_A$ for a matrix $A$ is $\myvar{y}^T A \myvar{y}$. A set $\myset{S}$ is forward invariant if for every $\myvar{x}(0) \in \myset{S}$, $\myvar{x}(t) \in \myset{S}$ $\forall t \in \myset{I}$, where $\myset{I}$ is the maximal interval of existence for the solution $\myvar{x}(t)$. A continuous function $\alpha:\mathbb{R} \to \mathbb{R}$ is an extended class-$\mathcal{K}$ function if it is strictly increasing and $\alpha(0) = 0$.

\section{Background}

\subsection{Control Barrier Functions}

Here we introduce existing work regarding control barrier functions for nonlinear affine systems: 
\begin{equation}\label{eq:nonlinear affine dynamics}
 \myvardot{x} = \myvar{f}(\myvar{x}) + \myvar{g}(\myvar{x}) \myvar{u}
\end{equation}
 where $\myvar{f}: \mathbb{R}^n \to \mathbb{R}^n$ and $\myvar{g}: \mathbb{R}^n \to \mathbb{R}^{n\times m}$, are locally Lipschitz functions, $\myvar{x}(t, \myvar{x}_0) \in \mathbb{R}^n$ is the state, $\myvar{u}(\myvar{x}(t)) \in \myset{U} \subseteq \mathbb{R}^m$ is the control input. Here we consider $\myset{U} = \mathbb{R}^m$. Without loss of generality, we consider the origin $\myvar{x} = 0$ as the desired equilibrium point of the system \eqref{eq:nonlinear affine dynamics}.

Let $h(\myvar{x}): \mathbb{R}^n \to \mathbb{R}$ be a continuously differentiable function, and let the associated constraint set be defined by:
\begin{equation} \label{eq:constraint set general}
\myset{C} = \{\myvar{x} \in \mathbb{R}^n: h(\myvar{x}) \geq 0\}, 
\end{equation} 

Safety is ensured by showing that the system states are always directed into the constraint set (Theorem 3.1 of \cite{Blanchini1999}). This condition is written as: $\dot{h}(\myvar{x}) \geq -\alpha(h(\myvar{x}))$ for an extended class-$\mathcal{K}$ function $\alpha$, and formally defined as:
\begin{definition}\label{def:zcbf}\cite{Ames2017}:
Given the set $\myset{C}$ defined by \eqref{eq:constraint set general} for a continuously differentiable function $h: \myset{E} \to \mathbb{R}$, the function $h$ is called a ZCBF defined on an open set $\myset{E}$ with $\myset{C} \subset  \myset{E} \subset \mathbb{R}^n$ if there exists a locally Lipschitz, extended class-$\mathcal{K}$ function $\alpha$ such that the following holds:
\begin{align}\label{eq:zcbf condition}
 \underset{\myvar{u} \in \myset{U}}{\text{sup}} [L_f h (\myvar{x}) + L_g h(\myvar{x}) \myvar{u} + \alpha(h(\myvar{x}))] \geq 0, \forall \myvar{x} \in \myset{E}
\end{align}
\end{definition}

If $h$ is a ZCBF, then \eqref{eq:zcbf condition} can be enforced, which is linear with respect to $\myvar{u}$. Many methods implement this condition in a quadratic program to define the constraint satisfying control $\myvar{u}$ \cite{Ames2019}. One example of such a controller is defined as follows:
\begin{align} \label{eq:consat qp ct}
\begin{split}
\myvar{u}(\myvar{x}) \hspace{0.1cm} = \hspace{0.1cm} & \underset{\myvar{u} \in \mathbb{R}^m}{\text{argmin}}
\hspace{.3cm} || \myvar{u} - \myvar{k}(\myvar{x}) ||_2^2   \\
& \text{s.t.} \hspace{0.8cm} L_f h(\myvar{x}) + L_g h(\myvar{x}) \myvar{u} \geq -\alpha(h(\myvar{x}))
\end{split}
\end{align}
where $\myvar{k}(\myvar{x}) \in \mathbb{R}^m$ is a locally Lipschitz control law. Many related methods construct barrier-function related control laws similar to \eqref{eq:consat qp ct}, but provide no guarantees of stability.

\subsection{Problem Formulation}

The objective is to develop a control law and specify a region of attraction that ensures  asymptotically stability and constraint satisfaction simultaneously. 
\begin{assumption}\label{asm:ZCBF exists}
There exists a continuously differentiable ZCBF $h:\myset{E} \to \mathbb{R}$ for the system \eqref{eq:nonlinear affine dynamics} where $\myset{C} \subset \myset{E} \subset \mathbb{R}^n$, $\myset{C}$ is defined by \eqref{eq:constraint set general},  $0 \in \rev{\mysetint{C}}$, and $\myset{E}$ is an open set.
\end{assumption}

\begin{assumption}\label{asm:LF exists}
There exists a locally Lipschitz control law $\myvar{k}: \myset{D} \to \mathbb{R}^m$ with $\myset{D} \subseteq \mathbb{R}^n$, $0 \in \myset{D}$ for which $\myvar{f}(0) + \myvar{g}(0) \myvar{k}(0) = 0$, and a continuously differentiable Lyapunov function $V:\myset{D} \to \mathbb{R}$, such that \eqref{eq:nonlinear affine dynamics} under $\myvar{u} =\myvar{k}(\myvar{x})$ is asymptotically stable\footnote{ see e.g., \cite{Khalil2002} for standard definitions} with respect to the origin. Furthermore if the given $V$ is non-increasing for the system \eqref{eq:nonlinear affine dynamics} under $\myvar{u} = \myvar{k}\rev{(\myvar{x})}$, then it is assumed that the conditions of LaSalle's principle (Corollary 4.1 of \cite{Khalil2002}) hold for \eqref{eq:nonlinear affine dynamics} under $\myvar{u} = \myvar{k}\rev{(\myvar{x})}$.
\end{assumption}

\begin{problem}\label{prob:main prob}
Consider the system \eqref{eq:nonlinear affine dynamics} and suppose Assumptions \ref{asm:ZCBF exists} and \ref{asm:LF exists} hold. Design a control law $\myvar{u}$ and define a region of attraction $\Gamma \subset \mathbb{R}^n$, $\Gamma \cap \myset{C} \neq \emptyset$, such that for all $\myvar{x}(0) \in \Gamma \cap \myset{C}$, $\myvar{x}(t) \in \myset{C}$ for all $t\geq 0$ and the system \eqref{eq:nonlinear affine dynamics} in closed-loop with $\myvar{u}$ is asymptotically stable with respect to the origin.
\end{problem}

\section{Proposed Solution}

Here we define a \textbf{CBF-stabilizable} level set to define the region of attraction given a ZCBF and a control law that asymptotically stabilizes \eqref{eq:nonlinear affine dynamics} with respect to the origin. We then define the proposed safe, stable control law $\myvar{u}^*$. Finally, we address a class of nonlinear systems for which the proposed method facilitates the design of the region of attraction. We further define a novel passive, safe, stabilizing controller for mechanical systems.

\subsection{Safe, Stabilizing Control Design}\label{ssec:main results}

We define a \textbf{CBF-stabilizable} set as follows:
\begin{definition}\label{def:cbf admissible set}
Consider the system \eqref{eq:nonlinear affine dynamics} and suppose Assumptions \ref{asm:ZCBF exists} and \ref{asm:LF exists} hold. Let $\Gamma_\nu:= \{\myvar{x} \in \myset{D}: V(\myvar{x}) \leq \nu\}$ for $\nu \in \mathbb{R}_{>0}$. For \rev{a given} locally Lipschitz continuous, positive-definite matrix $G(\myvar{x}) \in \mathbb{R}^{m\times m}$, \rev{$\Gamma_\nu$ is \textbf{CBF-stabilizable} if 
\begin{align}\label{eq:A set}
\nabla V^T(\myvar{x})\myvar{g}(\myvar{x}) G^{-1}(\myvar{x}) \myvar{g}(\myvar{x})^T \nabla h(\myvar{x}) < 0 
\end{align}
holds on $\Gamma_\nu \cap \myset{C}$ except on a subset $\myset{A} \subset \Gamma_\nu \cap \rev{\myset{C}}$ for which the following holds:}
\begin{equation}\label{eq:delta ball CBF admissible def}
\rev{z(\myvar{x}) := L_fh(\myvar{x}) + L_gh(\myvar{x}) \myvar{k}(\myvar{x})  + \alpha(h(\myvar{x})) \geq  0} , \forall \myvar{x} \in \myset{A}.
\end{equation}
\end{definition}

The intuition behind $\myset{A}$ is that in this set, the control barrier function condition ($\dot{h} =\nabla h(\myvar{x})^T \myvardot{x} \geq - \alpha(h)$) may act to de-stabilize the system. For stability, $\dot{V}$ must be negative definite such that $\nabla V(\myvar{x})^T \myvardot{x} \leq 0$. Thus the inner product of $\nabla V(\myvar{x})$ and $\nabla h(\myvar{x})$ dictates if the conditions for safety and stability may counteract one another. Note that by construction, this inner product is weighted by $\myvar{g}(\myvar{x}) G^{-1}(\myvar{x}) \myvar{g}(\myvar{x})^T$, for which the reason will become clear in the later analysis. It is important to emphasize the influence of $G(\myvar{x})$, which is a design parameter for the proposed control law (to be defined) and can be used to design $\myset{A}$ for compatibility. 

In the following Lemma, we ensure that Definition \ref{def:cbf admissible set} is well-defined in that there always exists a \textbf{CBF-stabilizable} set:
\begin{lemma}\label{lem:existenxe of CBF set}
Consider the system \eqref{eq:nonlinear affine dynamics} and suppose Assumptions \ref{asm:ZCBF exists} and \ref{asm:LF exists} hold. For any locally Lipschitz continuous, positive-definite matrix $G(\myvar{x}) \in \mathbb{R}^{m\times m}$, there exists a $\bar{\nu} \in  \mathbb{R}_{>0}$ such that $\Gamma_{\bar{\nu}}$ is \textbf{CBF-stabilizable}.
\end{lemma}
\begin{proof}
Let $\myset{G}$ be a compact neighborhood of the origin for which $0 \in \mysetint{G}$. Due to the fact that $0 \in \text{Int}(\myset{C})$, where $\myset{C}$ is given in \eqref{eq:constraint set general}, we restrict $\myset{G} \subset \myset{C}$ such that $h(\myvar{x}) > 0$ for all $\myvar{x} \in \myset{G}$. Let $\myvar{y}(\myvar{x}) = \myvar{f}(\myvar{x}) + \myvar{g}(\myvar{x}) \myvar{k}(\myvar{x})$. By Lipschitz continuity of $\myvar{y}(\myvar{x})$ and the fact that the origin is an equilibrium point, it follows that $||\myvar{y}(\myvar{x}) - 0 || \leq L || \myvar{x} - 0 || \leq L \| \myvar{x} \|$, where $L \in \mathbb{R}_{>0}$ is the Lipschitz constant of $\myvar{y}(\myvar{x})$ for all $ \myvar{x} \in \myset{G}$.

Due to continuity of $h$ and $\nabla h$ and the fact that $h(\myvar{x}) > 0 $ in $\myset{G}$, it follows that there exists a $\varepsilon \in \mathbb{R}_{>0}$ such that $\frac{\alpha(h(\myvar{x}))}{L ||\nabla h(\myvar{x})||} \geq \varepsilon$ for all $\myvar{x} \in \myset{G}$. Let $\delta\in (0, \varepsilon]$ be sufficiently small such that $\myset{B}_{\delta} \subseteq \myset{G} \subset \myset{C}$. Then it follows that $ || \myvar{x} || \leq \delta \leq \frac{\alpha(h(\myvar{x}))}{L ||\nabla h (\myvar{x})||} $ for all $\myvar{x} \in \myset{B}_\delta \subset \myset{C}$. Thus by construction $\| \myvar{f}(\myvar{x}) + \myvar{g}(\myvar{x}) \myvar{k}(\myvar{x}) \| \leq L \delta \leq  \frac{\alpha(h(\myvar{x}))}{||\nabla h (\myvar{x})||}$\rev{, which yields: $-\|\nabla h\| \|\myvar{f}(\myvar{x}) + \myvar{g}(\myvar{x}) \myvar{k}(\myvar{x})\| + \alpha(h(\myvar{x}))\geq 0$. By using the Cauchy-Schwartz inequality it follows that $z(\myvar{x}) \geq -\|\nabla h\| \|\myvar{f}(\myvar{x}) + \myvar{g}(\myvar{x}) \myvar{k}(\myvar{x})\| + \alpha(h(\myvar{x}))\geq 0$ }for all $\myvar{x} \in \myset{B}_\delta$. Now we choose $\bar{\nu}$ sufficiently small such that $\Gamma_{\bar{\nu}} \subset \myset{B}_{\delta}$. Note that we can always find a sufficiently small $\bar{\nu}$ because $\myset{B}_\delta$ is non-empty and $V(\myvar{x})$ is a continuous, positive definite function with $V(0) = 0$. Since \rev{$\Gamma_{\bar{\nu}} \subset \myset{B}_\delta \subset \myset{C}$, $z(\myvar{x}) \geq 0$ holds for all $\myvar{x} \in \Gamma_{\bar{\nu}}$ and so \eqref{eq:delta ball CBF admissible def} holds for any $\myset{A} \subset \Gamma_{\bar{\nu}} = \Gamma_{\bar{\nu}} \cap \myset{C}$.} Consequently $\Gamma_{\bar{\nu}}$ is \textbf{CBF-stabilizable}.
\end{proof}

\begin{remark}\label{rem:nontrivial CBF set}
We note that Lemma \ref{lem:existenxe of CBF set} ensures a \textbf{CBF-stabilizable} set always exists by shrinking $\Gamma_\nu$ until $\Gamma_\nu \subset \myset{C}$. However this is a conservative approach to ensuring safety and we aim to avoid this by addressing $\myset{A}$ in Definition \ref{def:cbf admissible set}. We must emphasize that the set $\myset{A}$ is a \emph{subset} of $\Gamma_\nu \cap \rev{\myset{C}}$ over which \eqref{eq:delta ball CBF admissible def} must hold. We do \emph{not} require that $\Gamma_\nu \subset \myset{C}$ as in Lemma \ref{lem:existenxe of CBF set}. On the contrary, we allow a designer to enlarge $\Gamma_\nu$ so long as the subset, $\myset{A}$, satisfies \eqref{eq:delta ball CBF admissible def}. This allows for larger regions of attraction and lets the system ``touch" the constraint boundary.
\end{remark}

Next we introduce the proposed safe, stabilizing control law:
\begin{subequations} \label{eq:proposed qp ct}
\begin{align} 
\myvar{u}(\myvar{x})^* \hspace{0.1cm} = \hspace{0.1cm} & \underset{\myvar{u} \in \mathbb{R}^m}{\text{argmin}}
\hspace{.3cm}  \frac{1}{2}\|\myvar{u} - \myvar{k}(\myvar{x})\|^2_{G(\myvar{x})} \label{eq:proposed qp cost}\\
& \text{s.t.} \hspace{0.8cm} L_f h(\myvar{x}) + L_g h(\myvar{x}) \myvar{u} \geq -\alpha(h(\myvar{x})) \label{eq:proposed qp constraint}
\end{align}
\end{subequations}
where $G(\myvar{x}) \in \mathbb{R}^{m\times m}$ is the locally Lipschitz continuous, positive-definite matrix from Definition \ref{def:cbf admissible set}. The controller \eqref{eq:proposed qp ct} seeks a $\myvar{u}$ that is ``minimally close" to $\myvar{k}(\myvar{x})$ to retain stability properties, while always respecting the ZCBF condition \eqref{eq:proposed qp constraint} to enforce forward invariance of $\myset{C}$. We note that \eqref{eq:proposed qp ct} is a generalized version of \eqref{eq:consat qp ct} (set $G = I_{n\times n}$)  and so the results presented here also apply to \eqref{eq:consat qp ct}.

We will make use of the following assumption to ensure local Lipschitz continuity of \eqref{eq:proposed qp ct}:
\begin{assumption}\label{asm:lipschitz}
Given a differentiable function $h: \myset{E} \to \mathbb{R}$ with the closed set $\myset{C} \subset \myset{E} \subseteq \mathbb{R}^n$ and a set $\myset{S} \subseteq \mathbb{R}^n$, suppose either of the following conditions hold for \eqref{eq:nonlinear affine dynamics}:
\begin{enumerate}
\item $L_g h(\myvar{x})\neq 0$ for all  $\myvar{x} \in \myset{S} \cap \myset{E}$; or
\item $L_g h(\myvar{x}) = 0$ only if $\myvar{x} \in \myset{S} \cap \mysetint{C} $, and if $L_g h(\myvar{x}) = 0$ then $\nabla h(\myvar{x}) = 0$ for all $\myvar{x} \in \myset{S} \cap \myset{E}$.
\end{enumerate}
\end{assumption}

Next we ensure the proposed control \eqref{eq:proposed qp ct} is well-defined:
\begin{thm}\label{thm:lip cont}
Consider the system \eqref{eq:nonlinear affine dynamics} and suppose Assumptions \ref{asm:ZCBF exists} and \ref{asm:LF exists} hold. Given any locally Lipschitz continuous, positive-definite matrix $G(\myvar{x}) \in \mathbb{R}^{m \times m}$ and a \textbf{CBF-stabilizable} set $\Gamma_\nu$, then $\forall \myvar{x} \in \myset{D} \cap \myset{E}$, the proposed control \eqref{eq:proposed qp ct} exists, is unique, and can be written in closed-form as:
\begin{align}\label{eq:control closed form}
\myvar{u}^*(\myvar{x}) = \begin{cases}
\barvar{u}(\myvar{x}), \text{ if } z(\myvar{x})< 0\\
\myvar{k}(\myvar{x}), \text{ if } z(\myvar{x}) \geq 0 
 \end{cases}
\end{align}
where
\begin{equation}\label{eq:ubar}
\barvar{u}(\myvar{x}):= \myvar{k}(\myvar{x}) -\frac{z(\myvar{x})}{|| L_g h(\myvar{x})^T ||_{G^{-1}}^2} G^{-1}(\myvar{x}) L_g h(\myvar{x})^T,
\end{equation}
\rev{for} $z(\myvar{x})$ \rev{from \eqref{eq:delta ball CBF admissible def}}. Furthermore, if Assumption \ref{asm:lipschitz} holds for the ZCBF $h$ and $\myset{S} = \Gamma_\nu$, then $\myvar{u}^*(\myvar{x})$ is locally Lipschitz continuous for all $\myvar{x} \in \Gamma_\nu \cap \myset{E}$.
\end{thm}
\begin{proof}
Positive-definiteness of $G(\myvar{x})$ ensures that \eqref{eq:proposed qp ct} is well-posed in that the cost function is positive definite and \eqref{eq:proposed qp constraint} is affine w.r.t $\myvar{u}$. This, along with Definition \ref{def:zcbf}, ensures there always exists a unique solution $\myvar{u}^*$ for all $\myvar{x} \in \myset{E}$ \cite{Nocedal2006}.

To derive $\myvar{u}^*$ in closed-form, we define the control action when the constraint in \eqref{eq:proposed qp constraint} is active (i.e holds with equality) by using the Schur-complement method from \cite{Nocedal2006} when $\|L_g h(\myvar{x})^T \|_{G^{-1}}^2 \neq 0$. We denote this action as $\barvar{u}$ defined in \eqref{eq:ubar}. Thus $\myvar{u}^* = \barvar{u}$ if \eqref{eq:proposed qp constraint} holds with equality, otherwise $\myvar{u}^* = \myvar{k}(\myvar{x})$, since $\myvar{k}(\myvar{x})$ is the optimal solution for the unconstrained version of \eqref{eq:proposed qp ct}. We similarly note if \eqref{eq:proposed qp constraint} holds with a strict inequality, then $\lambda^*(\myvar{x}) = 0$, where $\lambda^*(\myvar{x})$ denotes the Lagrange multiplier of \eqref{eq:proposed qp ct}.

To derive \eqref{eq:control closed form}, we claim that $\myvar{u}^* = \myvar{k}(\myvar{x})$ only if $z(\myvar{x}) \geq 0$. First, consider when $z(\myvar{x}) > 0$ such that $L_f h(\myvar{x}) + L_g h(\myvar{x}) \myvar{k}(\myvar{x}) > -\alpha(h(\myvar{x}))$. Then the minimizing solution of \eqref{eq:proposed qp cost} is $\myvar{u}^* = \myvar{k}(\myvar{x})$ as \eqref{eq:proposed qp constraint} holds with strict inequality. Now consider the case when $z(\myvar{x}) = 0$ such that \eqref{eq:proposed qp constraint} holds with equality. Substitution of $z(\myvar{x}) = 0$ into $\myvar{u}^*(\myvar{x}) = \barvar{u}(\myvar{x})$ yields $\myvar{u}^*(\myvar{x}) = \myvar{k}(\myvar{x})$. Finally, consider when $z(\myvar{x}) < 0$ which yields: $L_f h(\myvar{x}) + L_gh(\myvar{x}) \myvar{k}(\myvar{x}) < -\alpha(h(\myvar{x}))$. Clearly \eqref{eq:proposed qp constraint} is not satisfied with $\myvar{u}^* = \myvar{k}(\myvar{x})$ and so $\myvar{u}^* \neq \myvar{k}(\myvar{x})$. Thus $\myvar{u}^* = \myvar{k}(\myvar{x})$ only if $z(\myvar{x}) \geq 0$. This implies that $\myvar{u}^* \neq \myvar{k}(\myvar{x})$ if $z(\myvar{x}) < 0$, and the only other solution for $\myvar{u}^*$ is $\myvar{u}^* = \barvar{u}$, which yields \eqref{eq:control closed form}. Note that by Definition \ref{def:zcbf} when $L_g h(\myvar{x}) = 0$ the ZCBF condition  \eqref{eq:proposed qp constraint} must hold independently of $\myvar{u}$ (i.e. any choice of $\myvar{u}$ will suffice). Thus when $L_g h(\myvar{x}) = 0$, since any $\myvar{u}$ satisfies \eqref{eq:proposed qp constraint}, the quadratic program \eqref{eq:proposed qp ct} will output the solution that minimizes the cost \eqref{eq:proposed qp cost}, namely $\myvar{u}^* = \myvar{k}(\myvar{x})$. As stated, $\myvar{u}^* = \myvar{k}(\myvar{x})$ only when $z(\myvar{x}) \geq 0$ and so the control law \eqref{eq:control closed form} is well-defined when $z(\myvar{x}) = 0$ and $L_g h(\myvar{x}) = 0$. 

To ensure local Lipschitz continuity we show that \eqref{eq:proposed qp ct} satisfies the conditions of Theorem 3.1 of \cite{Hager1979}. The premise of this proof is to check that the control law is ``well-behaved" (i.e $\myvar{u}^* = \myvar{k}(\myvar{x})$) in a neighborhood around $L_g h(\myvar{x}) = 0$, which by assumption is restricted to $\mysetint{C} \cap \Gamma_\nu$ (for $\myvar{x} \in \Gamma_\nu \cap \myset{E}$) and implies that $\nabla h(\myvar{x}) = 0$. Then for any states outside of this neighborhood, $L_gh \neq 0$ and does not get arbitrarily close to $0$ such that the control is well-defined. To do so, we must first show that for every point $\barvar{x} \in \Gamma_\nu \cap \myset{E}$ for which $L_g h(\barvar{x}) = 0$, there exists a compact set $\myset{B}_\Delta(\barvar{x})$ such that $\myvar{u}^* = \myvar{k}(\myvar{x})$ for all $\myvar{x} \in \myset{B}_\Delta(\barvar{x})$. As mentioned in the previous paragraph, if $L_g h(\barvar{x})= 0$ then $\myvar{u}^* = \myvar{k}(\barvar{x})$. We choose a $\Delta' \in \mathbb{R}_{>0}$ such that $\myset{B}_{\Delta'}(\barvar{x}) \subset \text{Int}(\myset{C})$. By continuity of $h$, $\alpha$, $\myvar{f}$, $\myvar{g}$, and $\myvar{k}$, and \eqref{eq:constraint set general}, there exists a $\varepsilon_1$, $\varepsilon_2 \in \mathbb{R}_{> 0}$ such that for all $\myvar{x} \in \myset{B}_{\Delta'}(\barvar{x})$, $\alpha(h(\myvar{x})) \geq \varepsilon_1$ and $||\myvar{y}(\myvar{x})|| \leq \varepsilon_2$ for $\myvar{y}(\myvar{x}) = \myvar{f}(\myvar{x}) + \myvar{g}(\myvar{x}) \myvar{k}(\myvar{x})$. Now we design $\Delta \leq \Delta'$, where by continuity of $\nabla h$, there exists a $\mu \in \mathbb{R}_{>0}$ such that $\|\nabla h(\myvar{x})\| \leq \mu$ for all $\myvar{x} \in \myset{B}_{\Delta}(\barvar{x})$. Let $\Delta$ be sufficiently small such that $\mu < \frac{\varepsilon_1}{\varepsilon_2}$ for all $\myvar{x} \in \myset{B}_\Delta(\barvar{x})$. Substitution of $\mu < \frac{\varepsilon_1}{\varepsilon_2}$  with the previous bounds yields: $|| \nabla h^T \myvar{y}(\myvar{x}) || \leq ||\nabla h|| ||\myvar{y}(\myvar{x})|| \leq ||\nabla h|| \varepsilon_2 < \varepsilon_1$. It follows that $|| L_f h(\myvar{x}) + L_g h(\myvar{x}) \myvar{k}(\myvar{x})|| < \varepsilon_1 \leq \alpha(h(\myvar{x}))$ for all $\myvar{x} \in \myset{B}_\Delta(\barvar{x})$ and so the choice of $\myvar{u}^* = \myvar{k}(\myvar{x})$ satisfies \eqref{eq:proposed qp constraint} for $\myvar{x} \in \myset{B}_\Delta(\barvar{x})$. Since $\myvar{u}^* = \myvar{k}(\myvar{x})$ is the optimal solution and renders the constraint inactive, then $\myvar{u}^* = \myvar{k}(\myvar{x})$ in a neighborhood of $\barvar{x}$ for which $L_g h(\barvar{x}) = 0$. Note that when $\myvar{x} \in \myset{B}_\Delta(\barvar{x})$, $\lambda^*(\myvar{x}) = 0$ follows since \eqref{eq:proposed qp constraint} is inactive.

We next show that $\myvar{u}^*$ is locally Lipschitz continuous by addressing the conditions of Theorem 3.1 of \cite{Hager1979}. For a given $\myvar{x}$, we construct a convex, compact set $\myset{H} \subset \Gamma_\nu \cap \myset{E}$ such that $\myvar{x} \in \myset{H}$. Note that although $\Gamma_\nu \cap \myset{E}$ may not be convex, there always exists a convex, compact neighborhood of $\myvar{x}$ for any $\myvar{x} \in \Gamma_\nu \cap \myset{E}$. We construct $\myset{H}$ as follows. If $\myvar{x} \in \Gamma_\nu \cap \myset{E}$ is such that $L_g h(\myvar{x})= 0$, we define $\myset{H} = \myset{B}_\Delta(\myvar{x})$ for which $\myvar{u}^*(\myvar{x}) = \myvar{k}(\myvar{x})$ in $\myset{H}$. For any $\myvar{x}$ such that $L_g h(\myvar{x}) \neq 0$, we let $\myset{H}$ be any compact, convex subset of $\Gamma_\nu \cap \myset{E}$ for which $L_g h(\myvar{x}) \neq 0$ in $\myset{H}$. As discussed previously, there always exists a solution to \eqref{eq:proposed qp ct} for all $\myvar{x} \in (\Gamma_\nu \cap \myset{E}) \supset \myset{H}$ such that condition A.1 of Theorem 3.1 \cite{Hager1979} is satisfied. 

By assumption, $G(\myvar{x})$ is a positive-definite matrix and a continuous function of $\myvar{x}$ such that $G$ is bounded in $\myset{H}$. Continuity of $\nabla h$ ensures it is bounded in $\myset{H}$. Thus there exist $\gamma_1, \gamma_2 \in \mathbb{R}_{>0}$, $\gamma_1, \gamma_2 < \infty$ such that $||G|| \leq \gamma_1$, $||L_g h|| \leq \gamma_2$ for all $\myvar{x} \in \myset{H}$, which satisfies A.2 of \cite{Hager1979}. 

Positive-definiteness and continuity of $G$ for $\myvar{x} \in \myset{H}$ ensures there exists a $\nu \in \mathbb{R}_{>0}$ such that $\myvar{u}^T G \myvar{u} \geq \nu ||\myvar{u}||^2$ for all $\myvar{x} \in \myset{H}$. Next, we show there exists a $\beta \in \mathbb{R}_{>0}$ such that $||(L_g h(\myvar{x}))^T \lambda^*|| \geq \beta ||\lambda^*||$ for all $\lambda^*$, for all $\myvar{x} \in \myset{H}$. We note that when $\lambda^* = 0$, the condition follows trivially. Now consider the case when $\myset{H} = \myset{B}_{\Delta}(\myvar{x})$. By construction, it follows that for all $\myvar{x} \in \myset{B}_{\Delta}(\myvar{x})$, $\lambda^* = 0$, and the condition follows trivially. For any $\myset{H}$ where $L_g h(\myvar{x}) \neq 0$, since $L_g h(\myvar{x})$ is continuous on the compact set $\myset{H}$, then there exists a $\beta$ such that $\| L_g h(\myvar{x})^T \| \geq \beta$.  Furthermore since $\lambda^*$ is a scalar it follows that $||L_g h^T \lambda^*|| = |\lambda^*| ||L_g h^T|| \geq \beta |\lambda^*|$.  Thus the condition $||L_g h^T \lambda^*|| \geq \beta |\lambda^*|$ holds for all $\lambda^*$, and for all $\myvar{x} \in \myset{H}$ and condition A.3 of \cite{Hager1979} is satisfied. Thus from Theorem 3.1 of \cite{Hager1979}, $\myvar{u}^*$ is  Lipschitz continuous on every $\myset{H}$, and so is locally Lipschitz continuous on $\Gamma_\nu \cap \myset{E}$.
\end{proof}

The following theorem states the main result of this paper and ensures that the control law $\myvar{u}^*$ along with a \textbf{CBF-stabilizable} level set ensures compatibility:
\begin{thm}\label{thm:zcbf con AS}
Consider the system \eqref{eq:nonlinear affine dynamics} and suppose all the conditions of Theorem \ref{thm:lip cont} hold. Then the system \eqref{eq:nonlinear affine dynamics} under the proposed control \eqref{eq:proposed qp ct} satisfies: 
\begin{enumerate}
\item $\myvar{x}(t) \in \Gamma_\nu \cap \myset{C}$ for all $t \geq 0$.
\item The closed-loop system is asymptotically stable with respect to the origin.
\end{enumerate}
\end{thm}
\begin{proof}
Theorem \ref{thm:lip cont} ensures $\myvar{u}^*$ exists, is unique, and is locally Lipschitz continuous on $\Gamma_\nu \cap \myset{E}$. Furthermore $\myvar{u}^*$ can be written in closed form as defined in \eqref{eq:control closed form}, \eqref{eq:ubar}.

1) We first prove forward invariance of $\Gamma_\nu \cap \myset{C}$. Substitution of $\myvar{x} = 0$ in \eqref{eq:control closed form} and the fact that $\myvar{f}(0) + \myvar{g}(0) \myvar{k}(0) = 0$ by Assumption \ref{asm:LF exists} ensures that the origin is an equilibrium point of the closed-loop system. Furthermore, since $0 \in \myset{C}$ and $V(0) = 0$, then $\Gamma_\nu \cap \myset{C} \neq \emptyset$.

Next, we prove that the Lyapunov function, $V$, for the nominal system (\eqref{eq:nonlinear affine dynamics} with $\myvar{k}(\myvar{x})$) is a valid Lyapunov function for the closed-loop system (\eqref{eq:nonlinear affine dynamics} with \eqref{eq:proposed qp ct}) on $\Gamma_\nu \cap \rev{\myset{C}}$. It follows trivially that when $\myvar{u}^* = \myvar{k}(\myvar{x})$, then $\dot{V} = \nabla V(\myvar{x})^T \myvar{y}(\myvar{x})  \leq 0$ for $\myvar{y}(\myvar{x}) = \myvar{f}(\myvar{x}) + \myvar{g}(\myvar{x}) \myvar{k}(\myvar{x})$. Thus we need to check $\dot{V}$ for when $\myvar{u}^* = \barvar{u}$. \rev{Computation of $\dot{V} $ along with substitution of $\myvar{u}^* = \barvar{u}$ yields: $\dot{V}  =  \nabla V(\myvar{x})^T(\myvar{f}(\myvar{x}) + \myvar{g}(\myvar{x}) \barvar{u} ) =  \nabla V(\myvar{x})^T \Big( \myvar{y}(\myvar{x}) -\frac{z(\myvar{x})}{|| L_g h(\myvar{x})^T ||_{G^{-1}}^2} \myvar{g}(\myvar{x}) G(\myvar{x})^{-1} \myvar{g}(\myvar{x})^T  \nabla h(\myvar{x}) \Big)$, and ultimately:}
\begin{align}\label{eq:Lyapunov ND}
\dot{V}  \leq  
 -\frac{z(\myvar{x})}{|| L_g h(\myvar{x})^T ||_{G^{-1}}^2} \nabla V(\myvar{x})^T \myvar{g}(\myvar{x}) G(\myvar{x})^{-1}\myvar{g}(\myvar{x})^T \nabla h(\myvar{x}) 
\end{align}
\rev{Since $z(\myvar{x}) \geq 0$ implies the trivial case when $\myvar{u}^* = \myvar{k}(\myvar{x})$, we need only consider when $z(\myvar{x}) < 0$. By \eqref{eq:A set}, $\dot{V} < 0$ for any $\myvar{x} \in (\Gamma_\nu \cap \myset{C}) \setminus \myset{A}$. Now consider $\myvar{x} \in \myset{A}$. By} assumption, \eqref{eq:delta ball CBF admissible def} is satisfied such that $\myvar{u}^* = \myvar{k}(\myvar{x})$ \rev{and} $\dot{V}\leq 0$ \rev{$\forall \myvar{x} \in \myset{A}$. Thus} $\dot{V} \leq 0$ \rev{$\forall \myvar{x} \in \Gamma_\nu \cap \myset{C}$.}
 
The previous analysis shows that for $\myvar{x}(t) \in \Gamma_\nu \cap \rev{\myset{C}}$, $\dot{V} \leq 0$ under the control \eqref{eq:proposed qp ct}. Since $\dot{V} \leq 0$ for every sub-level set of $V$ in $\Gamma_\nu \cap \myset{C}$, we know that $\myvar{x}$ can not escape $\Gamma_\nu \cap \myset{C}$ via the boundary $\partial \Gamma_\nu$ (see proof of Theorem 4.1 of \cite{Khalil2002}). Furthermore, since $h$ is a ZCBF and $\dot{h}(\myvar{x}) \geq - \alpha(h(\myvar{x}))$, Brezis Theorem (Theorem 4 of \cite{Redheffer1972}) ensures that $\myvar{x}(t) \in \myset{C}$ for all $t \in [0, T)$, where $[0, T)$ is the maximal interval of existence of $\myvar{x}(t)$ for some $T \in \mathbb{R}_{>0}$. We want to show that $T = \infty$, and we prove this by repeated application of Theorem 3.1 of \cite{Khalil2002} and Brezis Theorem as follows. 

First, since $\myvar{x}(0) \in \Gamma_\nu \cap \myset{C}$ and the closed-loop system is locally Lipschitz on $\Gamma_\nu \cap \myset{E}$ (which contains $\Gamma_\nu \cap \myset{C}$) via Theorem \ref{thm:lip cont}, Theorem 3.1 of \cite{Khalil2002} ensures there exists a $t_1 \in \mathbb{R}_{>0}$ such that $\myvar{x}(t)$ is unique and exists on $[0, t_1]$. Now since all the conditions of Brezis Theorem hold on $[0, t_1]$, we know that $\myvar{x}(t) \in \Gamma_\nu \cap \myset{C}$ for $t \in [0, t_1]$. Since $\myvar{x}(t_1) \in \Gamma_\nu \cap \myset{C}$, we repeat this analysis for all $i \geq 2$ for which $t_i \in \mathbb{R}_{>0}$, $t_i > t_{i-1}$, such that $\myvar{x}(t)$ is uniquely defined on each $[0, t_i]$ and $\myvar{x}(t) \in \Gamma_\nu \cap \myset{C}$ for $t \in [0, t_i]$. 

Next we show that as $i \to \infty$ we have $t_i \to \infty$. Suppose instead that as $i \to \infty$, we have $t_i \to \tau$, $\tau < \infty$, then $\myvar{x}(t)$ would need to leave every compact subset of $\Gamma_\nu \cap \myset{E}$ (see proof of Theorem 3.3 of \cite{Khalil2002}). However, any solution that leaves $\Gamma_\nu \cap \myset{C}$ must traverse $(\Gamma_\nu \cap \myset{E}) \setminus (\Gamma_\nu \cap \myset{C})$ for which the closed-loop system is still locally Lipschitz and Brezi's theorem still holds. So those solutions could never have left $\Gamma_\nu \cap \myset{C}$, leading to a contradiction. Thus $\myvar{x}(t)$ is defined on $[0, \infty)$ and  $\myvar{x}(t) \in \Gamma_\nu \cap \myset{C}$ for all $t \geq 0$. We emphasize that since $\Gamma_\nu \cap \myset{C}$ is forward invariant we need only check Lipschitz continuity of the closed-loop system on $\Gamma_\nu \cap \myset{E}$, and hence we allow $\nabla h(\myvar{x}) = 0$ on $\mysetbound{C}$ for states outside of $\Gamma_\nu$.

2) If \eqref{eq:Lyapunov ND} holds with a strict inequality on $(\Gamma_\nu \cap \myset{C})\setminus \{0\}$, then asymptotic stability of the closed loop system under \eqref{eq:proposed qp ct} follows via Theorem 4.1 of \cite{Khalil2002}. For a non-increasing $V$, we must show that the only solution that can stay identically in $\{ \myvar{x} \in \mathbb{R}^n: \dot{V}(\myvar{x}) = 0\}$ is $\myvar{x}(t) \equiv 0$ to apply LaSalle's invariance principle (see Corollary 4.1 of \cite{Khalil2002}). We do this by showing that $\dot{V} = 0$ only occurs when $\myvar{u}^* = \myvar{k}(\myvar{x})$. We need to check when $\myvar{u}^* \neq \myvar{k}(\myvar{x})$ (i.e. $z(\myvar{x}) < 0$). First, when $\nabla V(\myvar{x})^T  \myvar{g}(\myvar{x}) G(\myvar{x})^{-1} \myvar{g}(\myvar{x})^T \nabla h(\myvar{x}) < 0$, then $\dot{V}$ is in fact negative definite and $\dot{V} \neq 0$. When $\nabla V(\myvar{x})^T  \myvar{g}(\myvar{x}) G(\myvar{x})^{-1} \myvar{g}(\myvar{x})^T \nabla h(\myvar{x}) \geq 0$ (i.e. $\myvar{x} \in \myset{A}$), then it follows that $\myvar{u}^* = \myvar{k}(\myvar{x})$. Thus $\dot{V} = 0$ only when $\myvar{u}^* = \myvar{k}$, and furthermore no new equilibrium points can arise. Now since the original system satisfies Corollary 4.1 of \cite{Khalil2002} by assumption, the same results follow, i.e., asymptotic stability of the closed loop system under \eqref{eq:proposed qp ct}.
\end{proof}

\begin{remark}\label{rem:non regular}
In previous work, $\nabla h \neq 0$ on $\mysetbound{C}$ was required \cite{Ames2019}. However, the conditions of Theorems \ref{thm:lip cont} and \ref{thm:zcbf con AS} are more relaxed in that $\nabla h = 0$ may occur on sections of $\mysetbound{C}$. The reason for this is that since $\myvar{u}^*(\myvar{x})$ keeps $\Gamma_\nu$ invariant, the states for which $\nabla h(\myvar{x}) = 0$ on $\mysetbound{C}$ will never be reached. 
\end{remark}

The ZCBF is used to improve the stability of the original nominal control law \rev{by increasing the rate at which the Lyapunov function decreases}. In this respect, \rev{if a designer can construct a \textbf{CBF-stabilizable} set, then they} need not find a new Lyapunov function to guarantee safety and stability. \rev{For passivity-based nominal controllers, the proposed methodology preserves passivity by ensuring a decrescent $V$, i.e., no additional energy is added to the system via the ZCBF controller.} This however means that the results presented herein are merely sufficient conditions for ensuring safety and stability. \rev{We further note that due to the use of a ZCBF and the results on asymptotic stability to the origin, the system \eqref{eq:nonlinear affine dynamics} in closed-loop with the proposed control is robust to bounded perturbations \cite{Xu2015a} \cite{Khalil2002}}.

\begin{remark}\label{rem:CLF-ZCBF}
Many existing methods require a CLF to define safe, stabilizing controllers \cite{Reis2021,Jankovic2018, Ames2017}. The assumption that a CLF is known is strong, but can be used in our controller \eqref{eq:proposed qp ct} by constructing a stabilizing control $\myvar{k}_{CLF}$  \cite{Sontag1989} to substitute for $\myvar{k}$ in \eqref{eq:proposed qp ct}.
\end{remark}

\subsection{A Special Class of Nonlinear Systems}\label{ssec:special systems}

The proposed methodology is dependent on computing the set $\myset{A}$, however this may be cumbersome as it depends on checking $\myvar{g}(\myvar{x})$ over all $\Gamma_\nu \cap \rev{\myset{C}}$. Here we exploit $G(\myvar{x})$ to facilitate the design process. We consider a class of nonlinear systems satisfying the following assumption:
\begin{assumption}\label{asm:full rank g}
The system \eqref{eq:nonlinear affine dynamics} can be written with $\myvar{g}(\myvar{x}) = E \barvar{g}(\myvar{x})$, where $E = [0_{(n-m)\times m}^T, I_{m \times m}]^T$, $m \in \{1,...,n\}$, and $\barvar{g}\in \mathbb{R}^{m \times m}$ is invertible\footnote{If $\myvar{g} = E \robvar{g}$ where $E = [0_{(n-l)\times m}^T, I_{l \times l}]^T$ and $\robvar{g} \in \mathbb{R}^{l\times m}$ with $l \leq m$ is full row rank, then we can satisfy this assumption by substituting $\myvar{u} = \robvar{g}^T \mu$ for which $\barvar{g} = \robvar{g} \robvar{g}^T$ and treat $\mu$ as the control input.} in a neighborhood of the origin.
\end{assumption}
Assumption \ref{asm:full rank g} applies to mechanical systems, for which constraint satisfaction and stability are highly desirable. For such systems, we propose the choice of $G(\myvar{x}) = \myvar{g}(\myvar{x})^T \myvar{g}(\myvar{x}) = \barvar{g}(\myvar{x})^T \barvar{g}(\myvar{x})$. Substitution into \eqref{eq:A set} yields: \rev{$\nabla V^T(\myvar{x}) E E^T \nabla h(\myvar{x}) < 0$}, which facilitates the design of \textbf{CBF-stabilizable} sets by completely removing the dynamic terms \rev{to check in \eqref{eq:A set}}. Now $\myset{A}$ is defined geometrically and can be checked without requiring the computation of $\myvar{g}(\myvar{x})$ \rev{nor $G(\myvar{x})$}. \rev{Next, we apply the proposed methodology to mechanical systems.}

Consider a mechanical system with generalized coordinates $\myvar{q} \in \myset{M} \subset \mathbb{R}^n$, generalized velocities $\myvar{v} \in \mathbb{R}^n$, and with the following dynamics:
\begin{equation}\label{eq:mechanical system dynamics}
\begin{split}
\myvardot{q} &= \myvar{v}  \\
\myvardot{v} &= M(\myvar{q})^{-1}(-C(\myvar{q}, \myvar{v}) \myvar{v}  - \myvar{\tau}_g(\myvar{q}) +  \myvar{u}  )
\end{split}
\end{equation}
where $M: \myset{M} \to \mathbb{R}^{n \times n}$ is the positive-definite inertia matrix, $C: \myset{M} \times \mathbb{R}^n \to \mathbb{R}^{n\times n}$ is the Coriolis and centrifugal matrix, and $\myvar{\tau}_g: \myset{M} \to \mathbb{R}^{n}$ is the generalized gravity, and  $\myvar{u} \in \mathbb{R}^n$. Let $(\myvar{q}(t, \myvar{q}_0), \myvar{v}(t,\myvar{v}_0)) \in \myset{M} \times \mathbb{R}^{n}$ be the solution of \eqref{eq:mechanical system dynamics} starting at $t = 0$, which for ease of notation is denoted by \rev{$\myvar{x} = (\myvar{q}, \myvar{v})$. The system \eqref{eq:mechanical system dynamics} satisfies Assumption \ref{asm:full rank g} with $E = [0_{n\times n}, I_{n\times n}]^T$.}

\subsubsection{Mechanical Systems with Relative-degree One ZCBF}\label{sssec:mech systems rel1}
\rev{Here, we apply the results to mechanical systems with relative-degree one ZCBFs:
\begin{proposition}\label{prop:mech sys rel1}
Consider the system \eqref{eq:mechanical system dynamics} with the nominal, computed torque control law $\myvar{u} = \myvar{k}(\myvar{x}) = C \myvar{v} + \myvar{g} + M(-K_p \myvar{q} - K_d \myvar{v})$ and Lyapunov function $V = \frac{1}{2}(\myvar{q}^T K_p \myvar{q} + \myvar{v}^T \myvar{v})$. Suppose Assumption \ref{asm:ZCBF exists} holds with $h = b - \frac{1}{2} (\myvar{q}^TP_q \myvar{q} +  \myvar{v}^TP_v \myvar{v})$, where $b \in \mathbb{R}_{>0}$,  $P_q \in \mathbb{R}^{n\times n}$ is a symmetric matrix, and $P_v \in \mathbb{R}^{n\times n}$ is a positive-definite, symmetric matrix. Then for $G = M^{-T} M^{-1}$, any $\Gamma_\nu \subset \myset{M} \times \mathcal{R}^n$ is \textbf{CBF-stabilizable}. Furthermore if Assumption \ref{asm:lipschitz} holds for the ZCBF $h$ and $\myset{S} = \Gamma_\nu$, then the results of Theorem \ref{thm:zcbf con AS} hold and \eqref{eq:mechanical system dynamics} in closed-loop with \eqref{eq:proposed qp ct} is passive with respect to the input term $\myvar{\mu} = -K_d \myvar{v}$ and output $\myvar{v}$. 
\end{proposition}
\begin{proof}
Assumption \ref{asm:LF exists} holds for the well-known computed torque control and associated $V$ with $\myset{D} = \myset{M} \times \mathbb{R}^n$ \cite{Murray1994}. Computation of the left-hand-side of \eqref{eq:A set} with the chosen $G$ (which is positive-definite and smooth \cite{Murray1994}) yields: $-\myvar{v}^TP_v \myvar{v}$, which is non-negative only when $\myvar{v} = 0$. Substitution of $\myvar{v} = 0$ into \eqref{eq:delta ball CBF admissible def} yields $z = \alpha(h) \geq 0$ in $\myset{A}$, such that any $\Gamma_\nu \subset \myset{D}$ is \textbf{CBF-stabilizable} according to Definition \ref{def:cbf admissible set}. If Assumption \ref{asm:lipschitz} holds, the results of Theorem \ref{thm:zcbf con AS} follow directly. For passivity, let $\myvar{y} = \myvar{v}$ and $V$ be the storage function such that substitution of $\myvar{\mu}$ into $\dot{V}$ for $\myvar{u} = \myvar{k}(\myvar{x})$ clearly yields $\dot{V} \leq \myvar{y}^T \myvar{\mu}$. Substitution of $\myvar{u} = \barvar{u}(\myvar{x})$ (for which $z < 0$) into $\dot{V}$ yields: $\dot{V} = \myvar{y}^T\myvar{\mu} + \frac{z}{\|L_g h^T\|_{G^{-1}}^2} \myvar{v}^TP_v \myvar{v} < \myvar{y}^T\myvar{\mu}$. Thus $\dot{V} \leq \myvar{y}^T\myvar{\mu}$ holds and the proof is complete.
\end{proof}
}

\subsubsection{Mechanical Systems with Relative-degree Two ZCBF}\label{sssec:mech systems}

\rev{We extend the results to the non-trivial case when the ZCBF is of higher order and to} the well-known control law $\myvar{k}(\myvar{x}) := \myvar{\tau}_g(\myvar{q}) -K_p \myvar{q} - K_d \myvar{v}$, for positive-definite, symmetric matrices $K_p, K_d \in \mathbb{R}^{n\times n}$, which asymptotically stabilizes \eqref{eq:mechanical system dynamics} to the origin with the following Lyapunov function:
\begin{equation}\label{eq:Lyapunov mechanical}
V(\myvar{x}) = \frac{1}{2} \myvar{v}^T M(\myvar{q}) \myvar{v} + \frac{1}{2}\myvar{q}^T K_p \myvar{q}
\end{equation}
\rev{We note that \eqref{eq:Lyapunov mechanical} is no longer quadratic as in the previous section, which complicates the design of a \textbf{CBF-stabilizable} set.} We will refer to the potential function as: $P(\myvar{q}) = \frac{1}{2} \myvar{q}^T K_p \myvar{q}$ with a level set $\myset{P}_\nu := \{\myvar{q} \in \myset{M}: P(\myvar{q}) \leq \nu \}$.

To construct a ZCBF, let $c: \myset{M} \to \mathbb{R}$ be a twice-continuously differentiable function with a locally Lipschitz Hessian. We define the safe set as:
\begin{equation}\label{eq:constraint set mech}
\myset{Q} := \{ \myvar{q} \in \myset{M}: c(q) \geq 0\}.
\end{equation}

Since $c(\myvar{q})$ is of relative-degree two with respect to \eqref{eq:mechanical system dynamics}, we construct our candidate ZCBF using a high-order barrier function approach \cite{Tan2021}:
\begin{equation}\label{eq:ZCBF mechanical}
h(\myvar{x}) = \dot{c}(\myvar{q}) + \phi(c(\myvar{q})) = \nabla c(\myvar{q})^T \myvar{v} + \phi(c(\myvar{q}))
\end{equation}
where $\phi$ is a \rev{continuously differentiable,} extended class-$\mathcal{K}$ function. We proceed by differentiating $h$ which yields $L_f h(\myvar{x})$ and $L_g h(\myvar{x})$ as follows:
\begin{align}\label{eq:lie derivatives mech}
L_fh(\myvar{x}) &=  \nabla c(\myvar{q})^T M(\myvar{q})^{-1} \left( -C(\myvar{q}, \myvar{v}) \myvar{v}  - \myvar{\tau}_g(\myvar{q}) \right) \\
&+ \frac{\partial \phi}{\partial c}\nabla c(\myvar{q})^T \myvar{v}  
+\myvar{v}^T \nabla^2 c(\myvar{q}) \myvar{v} \nonumber \\
L_g h(\myvar{x}) &= \nabla c(\myvar{q})^T M(\myvar{q})^{-1} \label{eq:lie derivatives mech g}
\end{align}

In the following Lemma, we ensure that $h$ is a zeroing control barrier function:
\begin{lemma}\label{lem:zcbf mech}
Consider the system \eqref{eq:mechanical system dynamics} with the sets $\mysetbar{C} = \{ \myvar{x} \in \myset{M} \times \mathbb{R}^n: h(\myvar{x}) \geq 0\}$ and $\myset{C}:= (\myset{Q} \times \mathbb{R}^n) \cap \mysetbar{C}$ for $h$ defined in \eqref{eq:ZCBF mechanical} and $\myset{Q}$ defined in \eqref{eq:constraint set mech}. If Assumption \ref{asm:lipschitz} holds for $h$ with a given open set $\myset{E} \supset \myset{C}$, $\myset{E} \subseteq \myset{M} \times \mathbb{R}^n$, and $\myset{S} = \myset{M} \times \mathbb{R}^n$, then there exists a controller $\myvar{u}$ such that \eqref{eq:zcbf condition} holds. Furthermore, if any $\myvar{u}$ satisfying \eqref{eq:zcbf condition} is locally Lipschitz continuous and implemented in closed-loop in \eqref{eq:mechanical system dynamics}, then $\myset{C}$ and $\myset{Q}$ are forward invariant with respect to $\eqref{eq:mechanical system dynamics}$.
\end{lemma}
\begin{proof}
Let $\alpha(h)$ be any extended class-$\mathcal{K}$ function. Consider the control: $\myvar{u}_{c} = M\left( C \myvar{v} + \myvar{\tau}_g + \frac{\nabla c}{||\nabla c||_2^2}( L_fh - \alpha(h)) \right) $ when $\nabla c \neq 0$. Since $M$ is positive-definite, $L_g h = 0$ if and only if $\nabla c = 0$, and by assumption $\nabla c = 0 $ implies that $\nabla h = 0$. When $\nabla c\neq 0$, then $\myvar{u}_c$ is well-defined and substitution of $\myvar{u}_c$ for $\myvar{u}$ in $\dot{h}$ yields: $L_fh + L_gh u_c = - \alpha(h)$. When $\nabla c = \nabla h = 0$, $\dot{h} = 0$, we choose any $\myvar{u} \in \mathbb{R}^n$. Since $\nabla h = 0$ only occurs in $\mysetint{C}$ (see Assumption \ref{asm:lipschitz}) and $\alpha$ is an extended class-$\mathcal{K}$ function, it follows that $\dot{h} = 0 > - \alpha(h)$. Thus there exists a $\myvar{u}$ satisfying \eqref{eq:zcbf condition}.

The function $h$ is in fact a high-order control barrier function as per Definition 5 of \cite{Tan2021} and so implementation of a locally Lipschitz $\myvar{u}$ satisfying \eqref{eq:zcbf condition} ensures $\myset{C}$ is forward invariant via Theorem 1 of \cite{Tan2021}. By construction, for any $\myvar{x}\in \myset{C}$, $\myvar{q} \in \myset{Q}$. Thus since $\myset{C}$ is forward invariant, so is $\myset{Q}$. 
\end{proof}

We proceed by constructing a $\myvar{u}^*$ similar to \eqref{eq:proposed qp ct} to safely stabilize the system. Substitution of \eqref{eq:proposed qp ct} into $\dot{V}$ for when $z  = L_f h + L_g h  \myvar{k} + \alpha (h)< 0$ yields: \rev{$\dot{V} = \frac{1}{2}\myvar{v}^T \dot{M} \myvar{v} + \myvar{v}^T ( - C \myvar{v}  + \myvar{\tau}_g + \myvar{u}^* )
 + \myvar{q}^T K_p \myvar{v}  = \myvar{v}^T ( - K_d \myvar{v} - z\frac{G^{-1} L_g h^T}{\| L_g h^T \|_{G^{-1}}^2} )  = - \myvar{v}^T  K_d \myvar{v} - z \frac{\myvar{v}^T G^{-1} M^{-1} \nabla c}{( \nabla c^T M^{-1} G^{-1} M^{-1} \nabla c )}$. We note that the skew-symmetric property of mechanical systems (i.e., $\myvar{v}^T (\frac{1}{2}\dot{M} - C) \myvar{v} = 0$) \cite{Murray1994} was used in this derivation.} Now we choose $G = M^{-1}$, which yields:
\begin{align}\label{eq:Vdot analysis mech}
\dot{V} &= - \myvar{v}^T  K_d \myvar{v} - z \frac{\dot{c}}{\| \nabla c\|_{M^{-1}}^2} 
\end{align}
and $\myset{A} = \{ \myvar{x} \in \Gamma_\nu \cap \rev{\myset{C}}: \dot{c} = \nabla c^T \myvar{v} \geq 0 \}$. From Theorem \ref{thm:zcbf con AS}, if we can find a $\Gamma_\nu$ for which $\dot{c} < 0$ whenever $z < 0$, then asymptotic stability follows. However it may be difficult to guarantee such a condition holds for general constraint sets. Here we deviate from the analysis of Section \ref{ssec:main results} and propose a novel, augmented form of $\myvar{u}^*$ to guarantee asymptotic stability (we drop dependencies for notational convenience):
\begin{align}\label{eq:control closed form aug}
\myvar{u}^*(\myvar{x}) = \begin{cases}
\myvar{k}, \text{ if } z \geq 0, \\
\barvar{u}, \text{ if } z < 0, \dot{c} \leq 0\\
\barvar{u} + \frac{\rho^2 z \dot{c}}{|| \nabla c ||_{M^{-1}}^2 } \myvar{v} , \text{ if } z < 0, \dot{c} > 0 
 \end{cases}
\end{align}
where $\barvar{u}  = \myvar{k} -\frac{z}{|| \nabla c ||_{M^{-1}}^2 } \nabla c$ and $\rho \in \mathbb{R}_{>0}$ is a gain parameter to be designed. The additional term,  $\myvar{\xi}:= \frac{\rho^2 z \dot{c}}{|| \nabla c ||_{M^{-1}}^2 } \myvar{v}$, in \eqref{eq:control closed form aug} cancels out the effects of $\dot{c}$ from \eqref{eq:Vdot analysis mech} for $\|\myvar{v} \|_2 > \frac{1}{\rho}$, and prevents the ZCBF condition from injecting additional energy into the system. With the new control law \eqref{eq:control closed form aug}, we can compute $\myset{A} = \{ \myvar{x} \in \Gamma_\nu \cap \rev{\myset{C}}: \dot{c} \geq 0, \|\myvar{v}\|_2 \leq \frac{1}{\rho} \}$. To address the case when the state enters $\myset{A}$, we present an alternative condition to \eqref{eq:delta ball CBF admissible def} for mechanical systems that is more practical to check a priori: 
\begin{align}\label{eq:condition for stability mech}
\psi(\myvar{q}):= - \nabla c^TM(\myvar{q})^{-1} K_p \myvar{q} + \alpha( \phi(c(\myvar{q}))) > 0, \forall \myvar{q} \in \myset{P}_\nu \cap \myset{Q}
\end{align}

Finally, if \eqref{eq:condition for stability mech} holds, we compute $\rho$ as follows:
\begin{align}\label{eq:rho tuning}
\rho = \max_{\myvar{q} \in \myset{P}_\nu \cap \myset{Q}} \frac{\eta_1(\myvar{q}) + \sqrt{\eta_1(\myvar{q})^2 + 4\psi(\myvar{q}) \eta_2(\myvar{q})}}{2\psi(\myvar{q})}
\end{align}
where $\eta_1(\myvar{q})$ $:= $ $k_c \|\nabla c^T M(\myvar{q})^{-1} \|_2$ $ + \| \nabla^2 c(\myvar{q}) \|_2$, $\eta_2(\myvar{q})$ $:= $ $\| \nabla c^T M(\myvar{q})^{-1} K_d \|_2$, and $k_c \in \mathbb{R}_{>0}$ is the bound on $C$ (i.e. $\|C\|_2 \leq k_c \|\myvar{v}\|_2$)\footnote{This is a well-known property of mechanical systems \cite{Spong1989}}. This tuning method only requires searching through the configuration space instead of the entire state space, which alleviates the computational complexity for implementation. Note that since $\psi > 0$ holds from \eqref{eq:condition for stability mech}, $\eta_1$, $\eta_2$, $\psi$ are continuous functions, and $\myset{P}_\nu \cap \myset{Q}$ is compact, $\rho$ is well-defined and bounded.

\begin{thm}\label{thm:mech safe as}
Consider the system \eqref{eq:mechanical system dynamics} and suppose the conditions of Lemma \ref{lem:zcbf mech} hold for $h$ defined by \eqref{eq:ZCBF mechanical}, $\myset{C}$ as defined in Lemma \ref{lem:zcbf mech}, and $\myset{Q}$ defined by \eqref{eq:constraint set mech}. If \eqref{eq:condition for stability mech} holds for a given $\myset{P}_\nu$, $\rho$ satisfies \eqref{eq:rho tuning}, and $\myvar{x}(0) \in \Gamma_\nu \cap \myset{C}$, then \eqref{eq:mechanical system dynamics} under $\myvar{u}^*$ from \eqref{eq:control closed form aug} ensures the following conditions hold:
\begin{enumerate}
\item $\myvar{q}(t) \in \myset{P}_\nu \cap \myset{Q}$ and $\myvar{x}(t) \in \Gamma_\nu \cap \myset{C}$, for all $t \geq 0$, 
\item The closed-loop system \eqref{eq:mechanical system dynamics} under \eqref{eq:control closed form aug} is asymptotically stable with respect to the origin,
\item The closed-loop system is passive\footnote{See \cite{Khalil2002} for standard definitions of passivity.} with respect to the input term $\myvar{\mu} = - K_d \myvar{v}$ in \eqref{eq:control closed form aug} and output $\myvar{v}$.
\end{enumerate}
\end{thm}
\begin{proof}
1) First, when $z \geq 0$ and $z < 0$, $\dot{c} \leq 0$ the control law is well-defined and locally Lipschitz from Theorem \ref{thm:lip cont}. Since \eqref{eq:condition for stability mech} holds, \eqref{eq:rho tuning} is well-defined and $\rho > 0$. Thus by using the same analysis as in Theorem \ref{thm:lip cont}, it follows that the additional term $\myvar{\xi}$ is also locally Lipschitz continuous on $\myset{E}$, where $\myset{E}$ is defined from Lemma \ref{lem:zcbf mech}. Thus \eqref{eq:control closed form aug} is well-defined in $\myset{E}$.

When $z \geq 0$ and $z<0, \dot{c} \leq 0$, we know that $\myvar{u}^*$ is equivalent to \eqref{eq:proposed qp ct} for the system \eqref{eq:mechanical system dynamics}. In this case it follows that $\dot{h} \geq - \alpha(h)$ which ensures $(\myvar{q}, \myvar{v}) \in \myset{C}$ as per Lemma \ref{lem:zcbf mech}. When $z<0$ and $\dot{c} > 0$, then $\myvar{u}^* = \barvar{u} + \myvar{\xi}$ for $\barvar{u}$ from \eqref{eq:ubar}. Implementation of $\myvar{u}^* = \barvar{u}$ ensures:  $\dot{h} = L_f h + L_g h \bar{u} \geq - \alpha(h)$. Thus implementation of $\myvar{u}^* = \barvar{u} + \myvar{\xi}$ yields: $\dot{h}  = L_f h + L_g h (\barvar{u} +  \myvar{\xi} )\geq - \alpha(h)  + L_g h \myvar{\xi}$, where $L_g h \myvar{\xi}$ is a perturbation. Here we show that this perturbation is benign in that forward invariance of $\myset{C}$ still holds. Let $\myset{I}:= [t_0, t_1) \subseteq \mathbb{R}_{\geq 0}$ denote the time interval over which $\myvar{\xi}$ is active. Since $\dot{h} \geq -\alpha(h)$ holds whenever $\myvar{\xi}$ is not implemented and by assumption $\myvar{x}(0) \in \myset{C}$, it follows that $c(\myvar{q}(t_0)) \geq 0$ and $h(\myvar{x}(t_0)) \geq 0$. Since $\myvar{\xi}$ is only implemented when $\dot{c} > 0$, it follows that $c \geq 0$ for all $t \in \myset{I}$, i.e., $\myvar{q} \in \myset{Q}$ on this interval. Furthermore, since both $\dot{c} > 0$ and $c \geq 0$ on $\myset{I}$, then $h \geq 0$ (see \eqref{eq:ZCBF mechanical}). Thus $\myvar{x}(t) \in \myset{C}$ on $\myset{I}$ and $\myvar{x}(t) \in \myset{C}$ for all $t \notin \myset{I}$ holds from Lemma \ref{lem:zcbf mech}. Thus $\myset{C}$ is forward invariant. We note again that by construction of $\myset{C}$, it follows that $\myset{Q}$ is also forward invariant.

From \eqref{eq:Vdot analysis mech}, it follows that \eqref{eq:mechanical system dynamics} under \eqref{eq:control closed form aug} ensures $\dot{V} \leq 0$ on $\Gamma_\nu \cap \myset{C}$ when (a) $z \geq 0$, (b) $z < 0$, $\dot{c} \leq 0$, and (c) $z < 0$, $\dot{c} > 0$, $\|\myvar{v} \|_2 > \frac{1}{\rho}$. Here we claim that the final possible case (d) (where  $z < 0$, $\dot{c} > 0$, $\|\myvar{v} \|_2 \leq \frac{1}{\rho}$) can never occur such that $\dot{V} \leq 0$ holds. We prove this claim by showing that for a slightly more general case than (d) where $\dot{c} \geq 0$, $\|\myvar{v}\|_2 \leq \frac{1}{\rho}$, $z$ must be non-negative. First, we use \eqref{eq:lie derivatives mech}, \eqref{eq:lie derivatives mech g}, and the definition of $z$ to separate $z$ as $z = \chi  + \theta$ where $\chi = -\nabla c^T M^{-1} C \myvar{v}  + \myvar{v}^T \nabla^2 c \  \myvar{v}- \nabla c^T M^{-1} K_d\myvar{v}$ and $\theta = -\nabla c^TM^{-1}K_p \myvar{q} + \frac{\partial \phi}{\partial c} \dot{c} + \alpha(\dot{c} + \phi(c))$. We can state that $\theta \geq \psi$ since $\dot{c}\geq 0$ and $\phi, \alpha$ are extended class-$\mathcal{K}$ functions. Now we can lower bound $\chi$ for $\| \myvar{v}\|_2 \leq \frac{1}{\rho}$ via $|\chi| \geq - \frac{1}{\rho}\eta_1(\myvar{q}) - \frac{1}{\rho^2} \eta_2(\myvar{q})$ such that $z \geq -  \frac{1}{\rho}\eta_1(\myvar{q}) - \frac{1}{\rho^2} \eta_2(\myvar{q}) + \psi $. Now $\rho$ as defined by \eqref{eq:rho tuning} is the solution to $\rho^2 \psi - \rho \eta_1 - \eta_2 = 0$, and since $\rho > 0$ it follows that $z \geq \psi -  \frac{1}{\rho}\eta_1(\myvar{q}) - \frac{1}{\rho^2} \eta_2(\myvar{q}) = 0$ for all $\myvar{q} \in \myset{P}_\nu \cap \myset{Q}$. Now from \eqref{eq:Lyapunov mechanical} and the fact that $M$ is positive definite, it follows that for all $\myvar{x} \in \Gamma_\nu \cap \myset{C}$, $\myvar{q}\in \myset{P}_\nu \cap \myset{Q}$ and so $z \geq 0$ holds in $\Gamma_\nu \cap \myset{C}$ for $\|\myvar{v}\|_2 \leq \frac{1}{\rho}$ and $\dot{c} \geq 0$. Thus the claim holds such that $\dot{V} \leq 0$ in $\Gamma_\nu \cap \myset{C}$, and so $\Gamma_\nu \cap \myset{C}$ is forward invariant. Since $\Gamma_\nu \cap \myset{C}$ is compact, it follows that $\myvar{x}(t) \in \Gamma_\nu \cap \myset{C}$ for all $t \geq 0$ using the same argument as in the proof of Theorem \ref{thm:zcbf con AS}. Since $\myvar{x} \in \Gamma_\nu \cap \myset{C}$ implies that $\myvar{q} \in \myset{P}_\nu \cap \myset{Q}$, then $\myvar{q}(t) \in \myset{P}_\nu \cap \myset{Q}$ for all $t\geq 0$ follows. Furthermore, since the above analysis holds when $\dot{c} = 0$, it follows that $\dot{V} = 0$ if and only if $\myvar{v} = 0$, and $\myvar{u}^* = \myvar{k}(\myvar{x})$ when $\myvar{v} = 0$.

2.) Let $\Omega = \{ \myvar{x} \in \Gamma_\nu \cap \myset{C}: \| \myvar{v} \| = 0\}$. Let $\myvar{v}(t)$ be a solution that belongs identically to  $\Omega$ i.e. $\myvar{v}(t) \equiv 0$ for all $t \geq 0$, for which $\myvardot{v}(t) \equiv \myvar{v}(t) \equiv 0$. Since $\myvar{u}^* = \myvar{k}(\myvar{x})$ when $\myvar{v} = 0$, the only solution that can stay identically in $\Omega$ is the origin, and the proof follows from Corollary 4.1 of \cite{Khalil2002}. 

3) Passivity follows by treating $V$ as the storage function and substituting $\myvar{\mu} = -K_d \myvar{v}$ and $\myvar{y} = \myvar{v}$ in $\dot{V}$. When $\myvar{u}^* = \myvar{k}(\myvar{x})$, it is clear that $\dot{V}= \myvar{y}^T \myvar{u}$. When $\myvar{u}^* = \barvar{u}(\myvar{x})$, then $\dot{V}= \myvar{y}^T \myvar{u} - z \frac{\dot{c}}{\|\nabla c\|_{M^{-1}}^2}$, and we know that $z< 0$, $\dot{c} \leq 0$ (see \eqref{eq:control closed form aug}) such that $\dot{V} \leq \myvar{y}^T \myvar{u}$. Finally, when $\myvar{u}^* = \barvar{u}(\myvar{x}) + \frac{\rho^2 z \dot{c}}{\|\nabla c\|_{M^{-1}}^2}$, then $\dot{V} =\myvar{y}^T \myvar{u} - \frac{z \dot{c}}{\|\nabla c\|_{M^{-1}}^2} (1- \rho^2 \|\myvar{v}\|_2^2)$. Since in this case $z < 0$, $\dot{c} > 0$ (see \eqref{eq:control closed form aug}) then $\dot{V} < \myvar{y}^T \myvar{u}$ for $\|\myvar{v} \|_2 > \frac{1}{\rho}$. Furthermore, from the previous analysis we need not check when $\|\myvar{v} \|_2 \leq \frac{1}{\rho}$ since this control  $\myvar{u}^* = \barvar{u}(\myvar{x}) + \frac{\rho^2 z \dot{c}}{\|\nabla c\|_{M^{-1}}^2}$ will never be active in this case. Thus $\dot{V} \leq \myvar{y}^T \myvar{\mu}$, which completes the proof.
\end{proof}

Theorem \ref{thm:mech safe as} does not require linearized approximations of the dynamics \cite{He2020} nor assumptions on Lipschitz continuity of the control, and it is not restricted to ellipsoidal sets \cite{Barbosa2020}. The condition \eqref{eq:condition for stability mech} can be checked offline and is only dependent on the coordinates $\myvar{q}$, \rev{yet this approach may still be subject to the curse of dimensionality as $n$ grows.}

\section{Numerical Examples}\label{sec:num examples}

\begin{figure}
     \centering
     \begin{subfigure}[b]{0.23\textwidth}
         \centering
         \includegraphics[width=\textwidth]{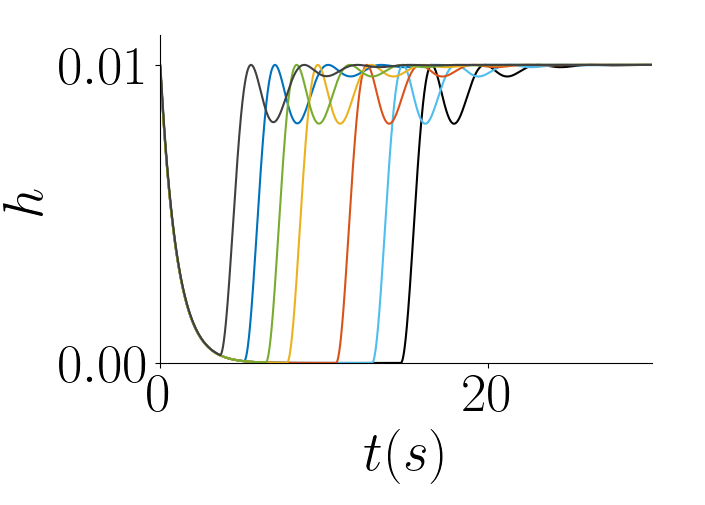}
         \caption{Relative-degree one ZCBF }
         \label{fig:Mech rel1 h}
     \end{subfigure}
     \hfill
     \begin{subfigure}[b]{0.245\textwidth}
         \centering
         \includegraphics[width=\textwidth]{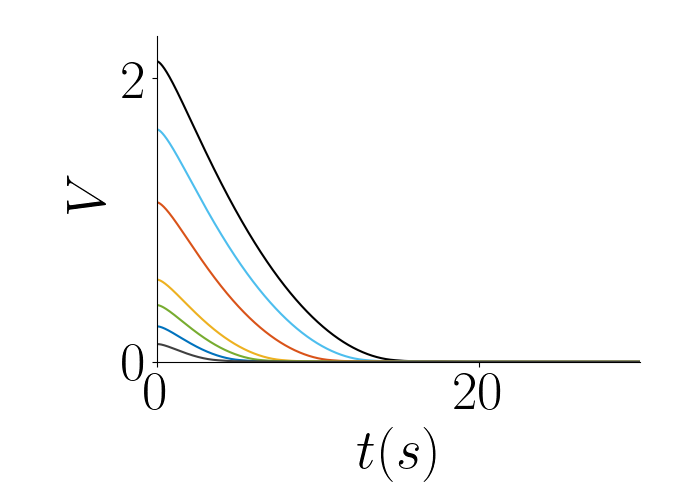}
         \caption{Lyapunov Function}
         \label{fig:Mech rel1 V}
     \end{subfigure}
        \caption{\rev{Trajectories resulting from the proposed control \eqref{eq:proposed qp ct} with $\myvar{k}(\myvar{x}) = C \myvar{v} + \myvar{g} + M(-K_p \myvar{q} - K_d \myvar{v})$.}}
        \label{fig:Mech rel1 results}
\end{figure}

Here we demonstrate the results on a 2-DOF manipulator consisting of two identical links with a length of $1$ m and mass of $1$ kg. We refer to \cite{Murray1994} for the system dynamics. \rev{First consider the relative-degree one candidate ZCBF with $P_q = 0_{n\times n}$, $P_v = I_{n\times n}$: $h(\myvar{x}) = b - \frac{1}{2} \|\myvar{v}\|_2^2$, which is used to restrict the squared-norm of the velocity of the closed-loop system below $b$. Now $h$ is a ZCBF because for any extended-class $\mathcal{K}$ function $\alpha$, the choice of $\myvar{u} = \myvar{g} + C\myvar{v} + \alpha(h) M \frac{v}{\|\myvar{v}\|_2^2}$, if $\|\myvar{v}\|_2 >0$, and $\myvar{u} = 0$, if $\|\myvar{v}\| = 0$, ensures that \eqref{eq:zcbf condition} holds. Furthermore $\nabla h = -\myvar{v} \neq 0$ on $\mysetbound{C}$. From Proposition \ref{prop:mech sys rel1}, the choice of $G = M^{-T} M^{-1}$ ensures that any $\Gamma_\nu$ is \textbf{CBF-stabilizable}. Figure \ref{fig:Mech rel1 results} shows the proposed control \eqref{eq:proposed qp ct} implemented for $\alpha(h) = h$, $K_p = I_{2\times 2}$, $K_d = 0.5 I_{2\times 2}$, and $b = 0.01$ for various initial conditions $\myvar{x}(0) \in \Gamma_\nu \cap \myset{C}$. We see that all trajectories converge to the origin, while ensuring $h \geq 0$, such that velocity constraints are always respected. }

Now consider the constraint set $\myset{Q}$ defined by \rev{the relative-degree two constraint} $\bar{c}(\myvar{q}) = a - (\myvar{q} - \myvar{q}_{r})^T P (\myvar{q} -  \myvar{q}_{r})$, for $a = 0.8$, $\myvar{q}_{r_1} = [0.9, 0]^T$, $P_1 = \text{diag}([1.0,2.0])$. The parameters of the stabilizing controller are $K_p = \text{diag}([1.0, 1.0])$, $K_d = \text{diag}([0.5, 0.5])$, with the ZCBF parameters as $\phi(h) = h$, $\alpha(h) = h$. We note that $\nabla \bar{c}$ = 0 at $\myvar{q}  = \myvar{q}_r$, which is a singularity where Assumption \ref{asm:lipschitz} would not hold. We remove the singularity using the technique from \cite{Tan2021} and define $c(\myvar{q}) = \chi_\delta(\bar{c}(\myvar{q}))$, where $\chi_\delta = 1$ if $\frac{\bar{c}}{\delta} > 1$ and $\chi_\delta = (\frac{\bar{c}}{\delta} - 1)^3 + 1$ if $\frac{\bar{c}}{\delta} \leq 1$. Since $c(\myvar{q}) = 0$ when $\bar{c}(\myvar{q})= 0$, the boundaries of their respective $\myset{Q}$ sets are the same. A value of $\delta = 0.7$ was used for these simulations. Furthermore, by construction when $\nabla c = 0$, we have $\nabla^2 c = 0$ such that the condition of Lemma \ref{lem:zcbf mech} is satisfied and $h$ defined by \eqref{eq:ZCBF mechanical} is a high-order ZCBF. For this choice of $P$ and $c$, \eqref{eq:condition for stability mech} holds for any $\myset{P}_\nu$. We note that \eqref{eq:condition for stability mech} was checked by performing a grid search.

\begin{figure}
     \centering
     \begin{subfigure}[b]{0.23\textwidth}
         \centering
         \includegraphics[width=\textwidth]{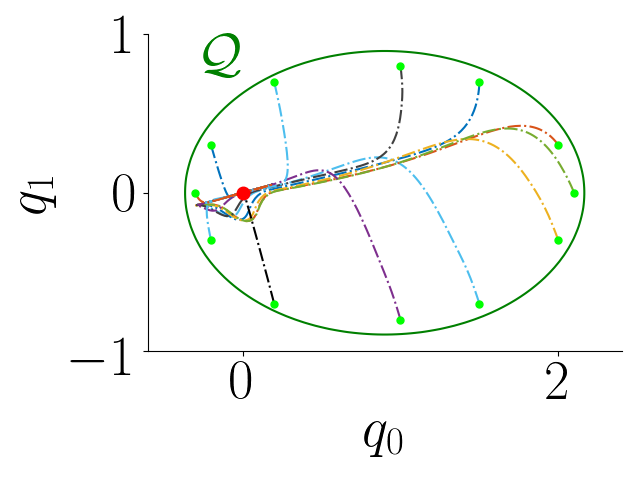}
         \caption{(unperturbed) }
         \label{fig:Mech phase}
     \end{subfigure}
     \hfill
     \begin{subfigure}[b]{0.245\textwidth}
         \centering
         \includegraphics[width=\textwidth]{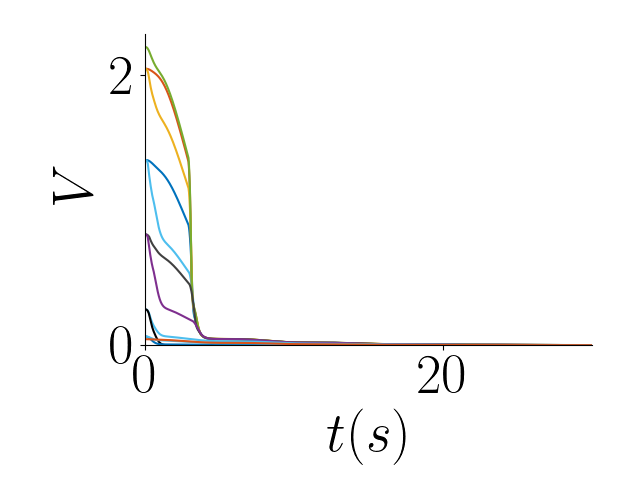}
         \caption{(unperturbed)}
         \label{fig:Mech V}
     \end{subfigure}
      \begin{subfigure}[b]{0.23\textwidth}
         \centering
         \includegraphics[width=\textwidth]{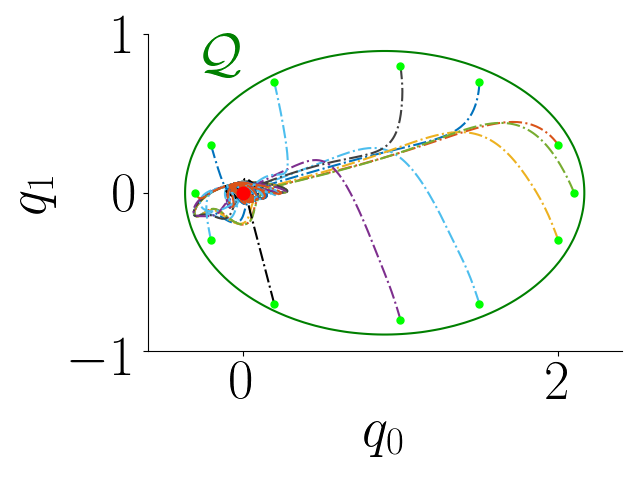}
         \caption{(perturbed) }
         \label{fig:Mech phase robust}
     \end{subfigure}
     \hfill
     \begin{subfigure}[b]{0.245\textwidth}
         \centering
         \includegraphics[width=\textwidth]{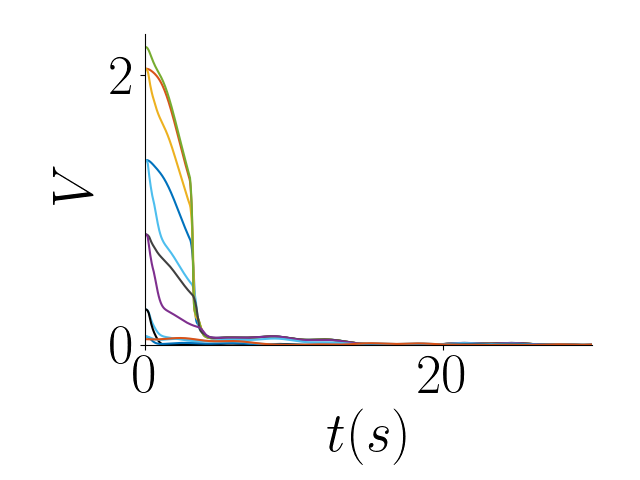}
         \caption{(perturbed)}
         \label{fig:Mech V robust}
     \end{subfigure}
        \caption{Phase plot and Lyapunov trajectories resulting from proposed control \eqref{eq:control closed form aug} for the 2-DOF manipulator without/with perturbations in a), b) and c), d), respectively. The dash-dotted curves represent the state trajectories $\myvar{q}(t)$ with initial points marked by a neon green star.}
        \label{fig:Mech results}
\end{figure}

Figure \ref{fig:Mech results} shows the resulting trajectories for initial conditions in $\myset{C}$ for the proposed control \eqref{eq:control closed form aug} applied to the 2-DOF manipulator. Figure \ref{fig:Mech phase} shows the trajectories remaining inside $\myset{Q}$ as they converge to the origin and Figure \ref{fig:Mech V} shows the resulting Lyapunov function decreasing to zero. Note that the Lyapunov function never increases, but does plateau in regions as expected due to the application of LaSalle's principle. To demonstrate robustness, we apply the time-varying perturbation $\myvar{d} = [d_0(t), d_1(t)]^T$ for $d_i = A_i \sin(\omega_i t)$ for $i \in \{0,1\}$ and $A_0 = A_1 = 0.1$, $\omega_0 = \omega_1 = 1.0$. The results of the closed-loop system with the proposed control \eqref{eq:control closed form aug} and matched disturbance $\myvar{d}(t)$ is shown in Figure \ref{fig:Mech results}, wherein the system remains bounded close to the origin and remains in $\myset{Q}$.

\section{Conclusion}

We proposed a methodology to ensure simultaneous stability and constraint satisfaction of nonlinear systems. The proposed approach defines a region of attraction, coined a \textbf{CBF-stabilizable} set, for which compatibility of an existing nominally stabilizing control law and ZCBF is ensured. A control law was proposed to ensure stability and safety in the \textbf{CBF-stabilizable set}. Furthermore, the proposed control was extended into a novel passive, safe, stabilizing controller for mechanical systems. Future work will incorporate input constraints in the proposed approach and \rev{investigate the relation between \textbf{CBF-stabilizable} sets and viability-kernels}.

\bibliographystyle{IEEEtran}
\bibliography{IEEEabrv,ShawCortez_CBFRH}

\end{document}